\documentclass[pre,twocolumn,superscriptaddress]{revtex4}

\usepackage{dcolumn}
\usepackage{graphicx}%
\usepackage{bm}%
\usepackage{amsmath}
\usepackage{amssymb} 

\begin{document}

\title{Polar Molecular Organisation in Liquid Crystals}
\author{A.G. Vanakaras}
\author{D.J. Photinos}
\affiliation{ \textit{Department of Materials Science, University of Patras, Patras 26500, Greece}}

\begin{abstract}

Various possibilities of polar self-organisation in low molar mass nematic, 
smectic and columnar liquid crystals are discussed with particular focus on 
the underlying molecular symmetries and interactions. Distinction is made 
between vector and pseudovector polarities, their quantification in terms of 
molecular order parameters and their relation to spontaneous electric 
polarisation and to molecular chirality. The understanding of the molecular 
mechanisms that give rise to polar ordering in existing lamellar and 
columnar phases may be useful for the design of new polar variants of common 
a-polar liquid crystals.

\textit{Keywords: }Polar Nematics; Ferroelectric Liquid Crystals; Polar Ordering.
\end{abstract}

\maketitle

\section{INTRODUCTION}

Phase anisotropy together with fluidity are the basic unifying features of 
all liquid crystals. Fluidity implies some disorder of the molecular 
positions in one or more spatial dimensions. Phase anisotropy reflects the 
directionality of the molecular organisation in these systems and can be of 
the a-polar type, if each direction in the bulk phase is physically 
indistinguishable from its opposite, or of the polar type, if such 
indistinguishability does not hold for all directions. In the latter case 
the underlying molecular organisation will exhibit polar orientational 
ordering. Accordingly, the bulk phase could in this case acquire spontaneous 
electric or magnetic polarisation if the molecules have permanent electric 
or magnetic dipoles. Aside from their purely scientific interest, the many 
actual and potential technological applications of liquid ferroelectric and 
ferromagnetic systems have stimulated intensive research on polar liquid 
crystals [1]. 

In this paper we discuss various molecular mechanisms that have been 
proposed for the understanding of polar ordering in the most common liquid 
crystalline phases. We consider low molar mass nematic, smectic and columnar 
liquid crystal phases and we concentrate on (i) the molecular interactions 
that could promote or oppose polar organisation and (ii) the quantification 
of polar orientational ordering in terms of order parameters and of 
macroscopic physical properties. Our considerations are restricted to 
electric properties and interactions but all the theoretical arguments can 
be carried over to magnetic systems as well.

\section{POLAR NEMATICS}

A polar nematic would be a translationally uniform phase with long-range 
polar orientational order. The most symmetric such phase is a uniaxial polar 
nematic, i.e. a nematic with full rotational symmetry about a unique 
phase-fixed axis denoted by the unit vector ${\bf N}$, the director, 
but with broken ``up-down'' symmetry so that the directions ${{\bf N}}$ 
and $-{{\bf N}}$ are not equivalent. The minimal molecular asymmetry 
requirements for the formation of such a phase correspond to uniaxially 
polar molecules, i.e. polar molecules with an axis ${{\bf m}}$ of full 
rotational symmetry, with the polarity rendering the molecule asymmetric 
with respect to the direction reversal ${{\bf m}} \to - {{\bf m}}$. 
Due to this asymmetry the ensemble average $\left\langle {{\bf m}} 
\right\rangle $ of the molecular unit vector ${{\bf m}}$ does not 
vanish. The single-molecule probability distribution is in this case a 
function of just one molecular variable, namely ${\rm {\bf m}} \cdot {\rm 
{\bf N}} = \cos \theta _{mN} $, where $\theta _{mN} $ is the angle of the 
molecular unit vector ${\rm {\bf m}}$ relative to the director ${\rm {\bf 
N}}$. The moments of this distribution are conveniently represented by the 
ensemble averages (order parameters) $\left\langle {P_l } \right\rangle 
\equiv \left\langle {P_l (\cos \theta _{mN} )} \right\rangle $ of the 
Legendre polynomials of rank $l$. The quantification of polar ordering for 
such a system is given by the ensemble average $\left\langle {\rm {\bf m}} 
\right\rangle = \left\langle {P_1 } \right\rangle {\rm {\bf N}}$. The 
primary measure of the magnitude of polarity is thus the first rank order 
parameter $\left\langle {P_1 } \right\rangle $. The second rank order 
parameter $\left\langle {P_2 } \right\rangle $, which is the leading order 
parameter for common (a-polar) nematics, is related to the breadth of the 
polar distribution according to the relation 
\begin{equation}
\label{eq1}
\left\langle {\cos ^2\theta _{mN} } \right\rangle - \left\langle {\cos 
\theta _{mN} } \right\rangle ^2 = (2 / 3)\left\langle {P_2 } \right\rangle - 
\left\langle {P_1 } \right\rangle ^2 + 1 / 3  ,
\end{equation}
\noindent
from which it becomes apparent that $\left\langle {P_1 } \right\rangle $ 
places the following lower bound on $\left\langle {P_2 } \right\rangle $,
\begin{equation}
\label{eq2}
\left\langle {P_2 } \right\rangle \ge (3 / 2)\left\langle {P_1 } 
\right\rangle ^2 - 1 / 2 \quad .
\end{equation}

If the molecules forming the uniaxial polar nematic phase carried permanent 
electric dipole moments ${{\bm \mu }}$ with a nonvanishing component 
$\mu _\parallel = {\rm {\bm \mu }} \cdot {\rm {\bf m}}$ along the direction 
of the molecular symmetry axis ${\rm {\bf m}}$, then the phase would exhibit 
spontaneous electric polarisation ${\rm {\bf P}}_s^e = {\cal N}\left\langle 
{\rm {\bm \mu }} \right\rangle $ where ${\cal N}$ denotes the molecular 
number density and $\left\langle {\rm {\bm \mu }} \right\rangle = \mu 
_\parallel \left\langle {\rm {\bf m}} \right\rangle $. Thus, the spontaneous 
polarisation vector would be in the direction of the phase symmetry axis 
${\rm {\bf N}}$ and proportional to the polar order parameter $\left\langle 
{P_1 } \right\rangle$ according to the relation
\begin{equation}
\label{eq3}
{\rm {\bf P}}_s^e = {\cal N}\mu _\parallel \left\langle {P_1 } \right\rangle 
{\rm {\bf N}} \quad .
\end{equation}
This equation relates, via a molecular property $\mu _\parallel $, a 
macroscopic electrostatic measure of polarity ${\rm {\bf P}}_s^e $ to a 
microscopic measure of polarity represented by the ``vector order 
parameter'' $\left\langle {\rm {\bf m}} \right\rangle = \left\langle {P_1 } 
\right\rangle {\rm {\bf N}}$. 

There does not seem to be any thermodynamic or symmetry argument forbidding 
\textit{a priory} the formation of a polar nematic phase, uniaxial or not. Yet, to date none 
of the existing low molar mass nematogens is known to exhibit such a 
phase[2]. It thus becomes theoretically challenging 
to rationalise, on the molecular level, the a-polarity of common nematics 
and from there to ask what sort of molecular interactions would be required 
to stabilise polar ordering in the nematic phase. For example, permanent 
electric dipole moments of a few Debyes are quite usual for molecules 
forming nematic phases and have measurable effects on their (a-polar) 
orientational order [3]. An obvious question is then 
whether electric dipole interactions could bring about phase polarity in 
nematics [4]. The answer is negative. A simplified 
explanation can be sketched as follows.

\begin{figure}[htbp]
\centerline{\includegraphics[width=3.4in]{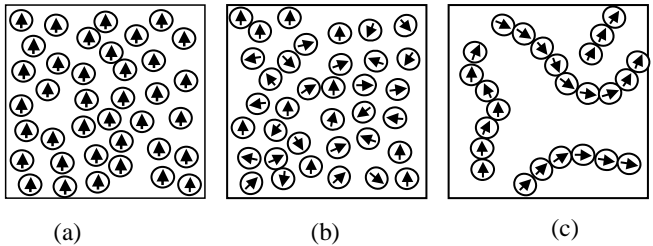}}
\caption{Figure 1 (a) Positionally disordered fluid of dipolar spheres in the 
perfectly oriented state and (b) in the orientationally disordered state. In 
both states the electrostatic contribution to the free energy of the system 
vanishes. (c) Molecular arrangements in the polar-string phase of the system 
at low densities.}
\label{fig1}
\end{figure}

The interaction potential between two dipoles, ${\rm {\bm \mu }}$ and ${\rm 
{\bm {\mu }'}}$ is given by 
\begin{equation}
\label{eq4}
u^{(d)} = [{\rm {\bm \mu }} \cdot {\rm {\bm {\mu }'}} - 3({\rm {\bm \mu }} 
\cdot {\rm {\bf r}})({\rm {\bm {\mu }'}} \cdot {\rm {\bf r}})] / r^3
\end{equation}
\noindent
where ${\rm {\bf r}}$ and $r$ are the interdipole unit vector and distance 
respectively. Consider the condensed, positionally disordered, phase of 
dipolar spherical molecules (i.e. molecules whose interactions in excess the 
dipole-dipole interaction are isotropic). If short-range 
position-orientation correlations are completely ignored then the 
contribution of the electrostatic interaction to the free energy for the 
phase with polar orientational order (Fig. 1a) vanishes. This is a 
reflection of the particular dependence of $u^{(d)}$ on ${\rm {\bf r}}$ for 
fixed dipole directions and interdipole distance. The most directly relevant 
feature of this dependence is illustrated in Fig. 2a, which shows that the 
energy of ``side-by-side'' configurations of parallel dipoles is half the 
magnitude and of opposite sign relative to the energy of the 
``head-to-tail'' configurations. Since in three dimensions there are two 
side-by-side configurations for each head-to-tail, the average dipole-dipole 
energy, in the absence of correlations, vanishes. The same null result is 
obtained when averaging the dipole-dipole contribution to the free energy in 
the orientationally disordered phase (Fig 1b), which however has larger 
entropy and will therefore always be thermodynamically more stable than the 
polarly ordered state. 

Essential to these arguments is the neglect of short-range correlations. The 
latter, however, are not always negligible. In fact, simulations of hard 
dipolar spheres have produced, for sufficiently strong dipoles and low 
densities, stable phases of polar strings, (linear aggregates of dipolar 
spheres in head to tail succession of the dipole moments, Fig. 1c) 
[5,6]. The correlations in 
these strings are such that the head-to-tail configuration dominates at 
short distances whereas the side-by-side (parallel or antiparallel) 
configurations correspond to molecular pairs on different strings, which are 
therefore more distant. This leads to an electrostatic reduction of the free 
energy that could compensate for the entropy reduction associated with the 
formation of the polar strings and thus provide thermodynamic stability 
relative to the completely disordered phase. On increasing the density, the 
side-by-side configurations are forced to the same average proximity as the 
head-to-tail and the chains eventually decompose into a positionally 
disorder a-polar fluid.

If instead of dipolar spherical molecules one considers elongated molecules 
with longitudinal dipoles then the polarly ordered condensed phase is 
further destabilised with respect to the a-polar phase. This happens because 
now, due to larger separations of the dipoles, the magnitude of the 
head-to-tail energy is less than twice the magnitude of the side by side 
energy (Fig. 2b) and therefore the positionally uncorrelated averaging will 
yield a net increase of the free energy of the polar phase relative to the 
a-polar one. Thus the stronger the dipole the more it disfavours the polar 
nematic ordering of the elongated molecules and in fact turns out to favour 
the a-polar smectic A phase[7]. 

\begin{figure}[htbp]
\centerline{\includegraphics[width=3.40in]{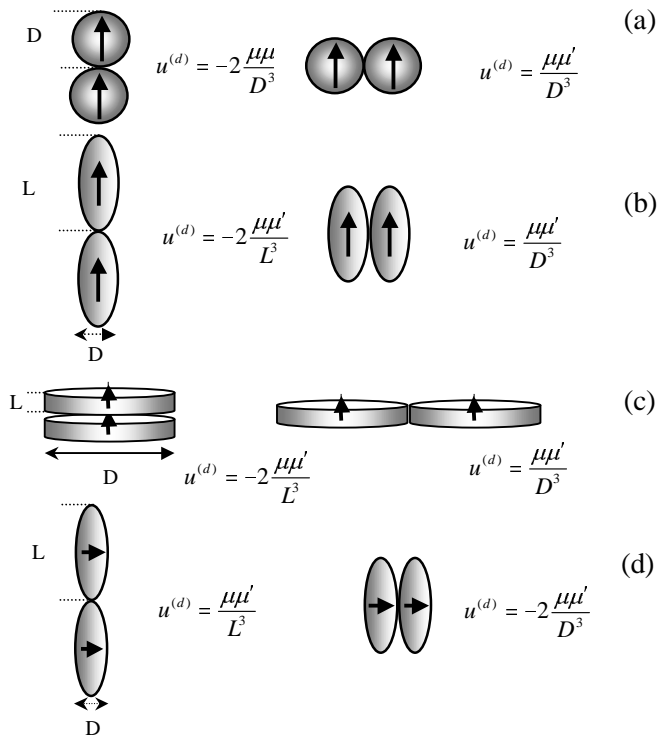}}
\caption{Dipole-dipole interaction energies for the parallel head-to-tail 
and side-by-side configurations of central dipoles carried by (a) spherical, 
(b) rod-like (longitudinally polar), (c) disc-like and (d) rod-like 
(transversely polar) molecules.}
\label{fig2}
\end{figure}

In the case of disc-like molecules with central dipoles along the disc 
symmetry axis, the electrostatic contribution favours the polar ordering. 
This is so because the head-to-tail energy, due to smaller interdipole 
separations, is now more than twice the magnitude of the side-by-side energy 
(see Fig. 2c). As shown in Monte Carlo (MC) simulations, however, the 
organisation of the polar discs does not correspond to a polar nematic but 
rather to a columnar phase with polar ordering within individual columns and 
antiferroelectric configurations of the columns [8].

It becomes apparent from these considerations that the formation of a polar 
nematic phase is not favoured by strong longitudinal electric dipoles. It 
rather requires interactions where the polar disposition of the molecules 
can be favoured but without forcing them to into any kind of positional 
register that would destroy the uniform distribution of their positions. 
Interactions of the amphiphilic type, for example, can favour the polar 
alignment of elongated molecules, as will be discussed below, but only as a 
result of (in the present context, at the expense of) phase microsegregation 
[9]. Furthermore, steric repulsions originating from 
the polar shape-asymmetry of tapered molecules, wedge shaped, pear shaped, 
etc, disfavours polar ordering in the undisturbed bulk phase since such 
shapes pack more efficiently in antiparallel configurations. It thus seems 
that none of the basic interactions encountered in common liquid crystals, 
nor any simple combination thereof, favours nematic polar ordering. This 
explains, at least partly, why a low molar mass nematic phase has not been 
detected experimentally to date. Recently, Berardi, Ricci and Zannoni 
[10] where successful in tailoring a model potential 
that can produce, under certain parameterisation, a polar nematic phase in 
MC simulations. Their model interaction essentially behaves in the opposite 
way to the dipole-dipole interaction (Fig. 2b) in that it assigns the lowest 
energy to the side-by-side parallel arrangement of a molecular pair, high 
energy to the parallel head-to-tail and intermediate energy to the 
side-by-side antiparallel. Of course the identification of real molecules 
that would interact in the required way is the major step to be taken. In 
the present state of molecular design efforts, however, it appears that much 
more is known on what should be avoided than on what is to be pursued. 

\section{ORTHOGONAL SMECTICS WITH POLAR LAYERS}

Polar ordering in orthogonal smectics is quite usual 
[11]. Known variants of the smectic A phase (\textit{Sm-A}) with 
polar layers include the bilayer phases $A_{d}$, $A_{2}$ and \textit{$\Gamma $} (Fig. 3a,b,c). 
In these phases the polar asymmetry is exhibited in the direction of the 
layer normal, with adjacent layers having opposite polarities thus rendering 
the bulk phase macroscopically a-polar. 

On the molecular scale, the structure of the phases $A_{2}$ and \textit{$\Gamma$} suggests 
that the molecular interactions giving rise to the polar ordering should be 
such as to favour the side-by-side parallel alignment of molecules within 
the same layer while favouring the head-to-head or tail-to-tail alignment of 
molecules in consecutive layers (Fig. 3a,c). Obviously neither the electric 
dipole forces nor the purely steric forces associated with wedge-type shape 
asymmetry favour such configurations; in fact they strongly disfavour them. 
In contrast, amphipile type forces favour both types of configurations and 
are therefore considered the primary interaction underlying the formation of 
these phases. 

The bilayer structure of the $A_{d}$ phase suggests molecular interdigitation 
among adjacent sublayers such that the favoured molecular arrangements 
produce side-by-side antiparallel configurations of the molecular tips (Fig. 
3b). These configurations could be favoured by interactions from dipole 
moments situated near the ends of the elongated molecules and in fact theory 
and simulations show that model rod-like molecules with off centre axial 
dipole moments can produce the $A_{d}$ phase structure 
[12]. However, the amphiphilic interactions also 
favour this type of configurations. Moreover, this arrangement makes the 
packing more effective in the presence of wedge like shape asymmetry, 
particularly in combination with amphipilicity. This is then an example of 
ordering where several types of interactions could separately or 
cooperatively produce the polarity. In fact, Photinos and Saupe 
[13] showed nearly three decades ago the theoretical 
possibility of longitudinally polar smectics based on a generic molecular 
field description of the polar interactions.

The type of polarity described above for orthogonal smectics is similar to 
that of nematics in that it consists of the local (within a sublayer) 
breaking of the up-down symmetry ${\rm {\bf N}} \Leftrightarrow - {\rm {\bf 
N}}$ along the director and entails the same minimal asymmetry of the 
molecular structure, i.e. uniaxial molecules with a polar axis ${\rm {\bf 
m}}$ of full rotational symmetry. Consequently, the polarity in each 
sublayer is quantified by the vector order parameter $\left\langle {\rm {\bf 
m}} \right\rangle$ as in uniaxial polar nematics. In the presence of 
longitudinal dipoles a spontaneous electric polarisation will be produced in 
each sublayer according to Eq. (\ref{eq3}) and, due to the alternating polarity of 
the bilayer structure, the resulting ``antiferroelectric'' system will 
appear electrically a-polar on the bilayer scale. The term longitudinal will 
be reserved for the above type of polarity, namely the one that manifests 
itself as the breaking of the ${\rm {\bf N}} \Leftrightarrow - {\rm {\bf 
N}}$ symmetry along the director (coincident with the layer normal in the 
case of orthogonal smectics), in order to distinguish it from the transverse 
polarity that is manifested as the breaking of the twofold rotational 
symmetry about the director \textbf{N}. 

\begin{figure}[htbp]
\centerline{\includegraphics[width=3.00in]{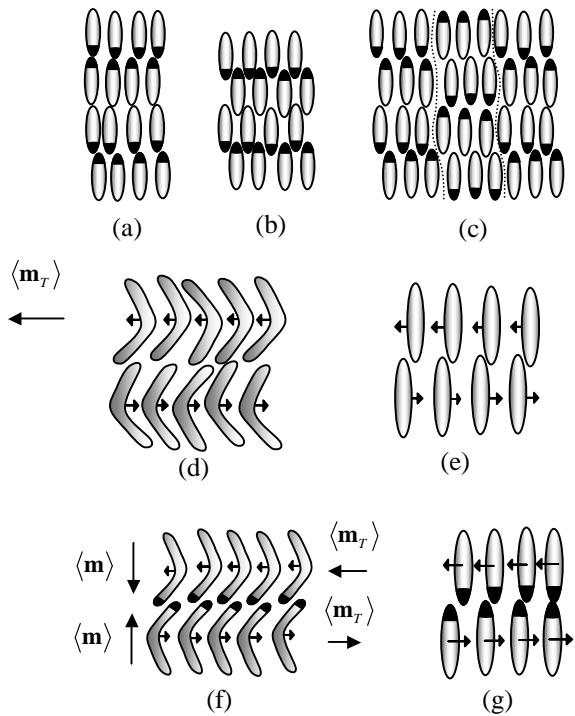}}
\caption{Schematic representation of molecular arrangements in some polar  
orthogonal smectics. (a) Bilayer \textit{Sm-A}$_{2.}$ (b) Semibilayer 
\textit{Sm-A}$_{d}$. (c) Ribbon bilayer \textit{Sm-$\Gamma $}. 
(d) Banana-type \textit{Sm-A}$_{P}$, shown here in antiferroelectric succession 
of the layers. (e) Rod-like polar smectic A phase model, shown here in 
antiferroelectric succession of the layers. (f) Banana-type orthogonal 
smectic phase model with both, longitudinal (along the layer normal) and 
transverse polarity, shown here in antiferroelectric succession of the 
layers with respect to both polarities. (g) Rod-like orthogonal smectic 
phase model with both, longitudinal (along the layer normal) and transverse 
polarity, shown here in antiferroelectric succession of the layers with 
respect to both polarities. The arrows attached to the molecules in figures 
(d) to (g) indicate the molecular vectors ${\rm {\bf m}}_T $.}
\label{fig3}
\end{figure}

An orthogonal smectic phase with transverse polarity and positional disorder 
has been reported recently [14] for compounds with 
banana-type molecular shape asymmetry. The polarity of the layers in this 
case is thought to result from the stacking of the molecules so that the 
``steric dipoles'' associated with their bent shape become aligned (see Fig. 
3d). This polar stacking can produce a spontaneous polarisation in the plane 
of the layers if the molecules possess electric dipole moments with 
nonvanishing component along the direction of the steric dipole. It should 
be noted that such electric dipole moments, if present, would contribute to 
the polar alignment because their configuration within each layer gives on 
average equal numbers of parallel side-by-side and head-to-tail pairs while 
the side-by-side pairs belonging to different layers are more distant and 
thus an overall reduction of the internal electrostatic energy relative to 
the a-polar system would result. In fact, unlike the case of longitudinal 
polarity, transverse polarity can in principle be generated by electric 
dipole interactions alone. Consider for example idealised smectic layers 
consisting of a planar arrangement of positionally disordered parallel 
rod--like molecules with central transverse electric dipoles (Fig. 3e). The 
internal electrostatic energy per molecule for the state with perfect 
alignment of the dipoles is lower than that of the state with randomly 
oriented dipoles by $\pi (\sigma / D)\mu ^2$, where $\mu$ is the dipole 
strength, $\sigma $ is the surface density of the molecules in the layers 
and $D$ is the diameter of the rods (the distance of closest approach of the 
dipoles, Fig 2d). The entropy per molecule for the polar state is lower than 
that of the random state by $k_B T\ln 2\pi$ and therefore the polar state 
becomes the thermodynamically stable one for sufficiently large values of 
the quantity $(\sigma / D)\mu^2 / k_B T$. At this level of approximation 
the intra-layer electrostatic interactions do not contribute to the internal 
energy of the system since the interaction energy between any dipole and all 
the dipoles belonging to a different layer averages to zero, both in the 
polar and the random states. This result suggests that, to a first 
approximation (i.e. ignoring orientational and positional fluctuations and 
correlations of inter- and intra-layer molecular motions), the inter-layer 
electrostatic interactions are indifferent to the mode of propagation of 
polarity across the layers (parallel or antiparallel disposition of adjacent 
layers). Monte Carlo simulations of rod like molecules with transverse 
dipoles give some indications of in-plane polar order 
[15]. Gil-Villegas, McGrother and Jackson 
[16] found that, even at fairly high densities, the 
dipoles in a layer self-organise to form ring domains that evolve, on 
lowering the temperature, into antiferroelectric elongated chain domains. 

It is apparent from the examples of banana shaped molecules and of rods with 
transverse dipoles in Fig. 3d,e that the minimal molecular asymmetry 
required for in-plane phase polarity corresponds to molecules with a plane 
of symmetry and a twofold symmetry axis ${\rm {\bf m}}_T $ on that plane. 
The molecular ``long axis'' ${\rm {\bf m}}$ is defined as the axis within 
the plane of symmetry and perpendicular to ${\rm {\bf m}}_T $ and is no 
longer an axis of rotational symmetry. The molecular polarity consists in 
the lack of ${\rm {\bf m}}_T \Leftrightarrow - {\rm {\bf m}}_T $ symmetry. 
Accordingly, phase polarity can be quantified by the vector order parameter 
$\left\langle {{\rm {\bf m}}_T } \right\rangle $. The direction of 
$\left\langle {{\rm {\bf m}}_T } \right\rangle $ defines a second director 
${\rm {\bf N}}_T $ in the plane of the layer and thus perpendicular to the 
primary director ${\rm {\bf N}}$. Consequently, transverse polarity makes 
the orthogonal smectic phase necessarily biaxial. The asymmetry associated 
with in-plane polarity consists in the breaking of the ${\rm {\bf N}}_T 
\Leftrightarrow - {\rm {\bf N}}_T $ symmetry while maintaining the ${\rm 
{\bf N}} \Leftrightarrow - {\rm {\bf N}}$ symmetry. When both symmetries are 
valid one has a biaxial (non-polar) orthogonal phase whereas the 
simultaneous breaking of both symmetries yields an orthogonal smectic with 
both in-plane and longitudinal polarity (see Fig. 3f,g). The types of 
molecular interactions that can produce the combined polarity can in 
principle be furnished by the superposition of the interactions giving rise 
to the individual polarities, for example amphiphilic interactions for the 
longitudinal polarity and banana type steric dipole for the transverse 
polarity (Fig. 3f), since such interactions are not incompatible or 
counteracting. It thus can be seen that the consideration of the two types 
of polar ordering and of their combination broadens the variety of 
orthogonal smectics with positionally disordered layers. These possibilities 
of polarity are carried over and further enriched for orthogonal smectics 
with in-plane positional order, the discussion of which is beyond the 
present scope.

\section{TILTED SMECTICS}

The least ordered of the tilted smectic phases is the \textit{Sm-C} phase. It consists of 
positionally disordered layers in which the orientational order of the 
molecules defines a director ${\rm {\bf N}}$ forming an angle (the tilt 
angle) with the direction normal to the layers. The \textit{Sm-C} layers have a plane of 
symmetry (the tilt plane, formed by the layer normal \textbf{Z} and the 
director \textbf{N}). They also have a twofold symmetry axis $C_{2}$ 
perpendicular to the plane of symmetry and a centre of inversion at the 
intersection of the twofold axis with the symmetry plane. The directions 
${\rm {\bf N}}$ and $ - {\rm {\bf N}}$ are equivalent (physically 
indistinguishable). However, the structure of the \textit{Sm-C} phase is inherently polar 
since the tilt singles out a unique direction about the layer normal. This 
unique direction is often represented by the so-called \textbf{C}-director 
[17] whose direction is defined from the projection 
of the \textbf{N}-director onto the layer plane but with ${\rm {\bf C}}$ and 
$- {\rm {\bf C}}$ describing physically distinct states. A particularly 
useful alternative [18] representation is provided 
by the tilt pseudovector \textbf{t} defined in terms of the layer normal 
\textbf{Z} and the \textbf{N}-director according to the relation
\begin{equation}
\label{eq5}
{\rm {\bf t}} = ({\rm {\bf Z}}\times {\rm {\bf N}})({\rm {\bf Z}} \cdot {\rm 
{\bf N}}) \quad .
\end{equation}
Obviously \textbf{t} is perpendicular to the tilt plane, it is invariant 
with respect to the replacement of \textbf{N} by --\textbf{N} and the states 
\textbf{t} and --\textbf{t} describe layers of opposite tilt and are 
therefore physically distinct. 
\begin{figure}[h]
\centerline{\includegraphics[width=3.1in]{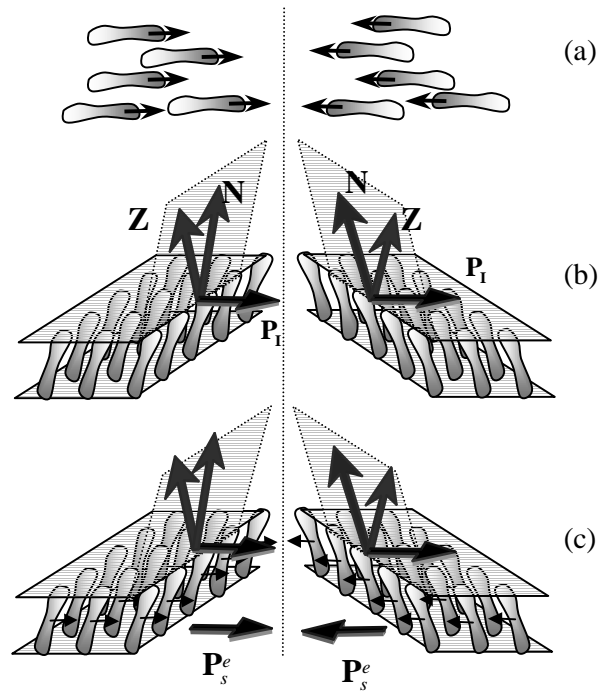}}
\caption{Molecular arrangements and their images in a mirror plane (dotted 
vertical line) perpendicular to the polar direction. (a) Polar nematic 
arrangement. Vector polarity breaks mirror symmetry; the two mirror images 
are distinct. (b) Tilted smectic layer, with the tilt plane containing the 
layer normal \textbf{Z} and the director \textbf{N}. Tilt-driven 
pseudovector polarity ${\rm {\bf P}}_{\rm {\bf I}} $ is compatible with 
mirror symmetry of the layer; the two mirror images are identical. (c) 
Tilted smectic layer with chiral molecules carrying transverse electric 
dipole moments. The resulting spontaneous electric polarization ${\rm {\bf 
P}}_s^e $ breaks mirror symmetry in the layer; the two mirror images are 
distinct.}
\label{fig4}
\end{figure}

The polarity of the \textit{Sm-C} phase, being clearly a direct consequence of the tilted 
ordering, is referred to as indigenous polarity 
[19]. It is of fundamentally different nature from 
the polarity that could appear in orthogonal smectics or in nematics. First, 
when polar ordering appears in orthogonal smectics or in nematics, mirror 
symmetry perpendicular to the polar director (see Fig. 4a) is broken. This 
type of polar ordering is referred to as vector-type polarity, in accordance 
with the sign reversal of vectors under mirror reflection. In contrast, the 
indigenous polarity of the \textit{Sm-C} phase is compatible with mirror symmetry in the 
plane normal to the direction of the polar asymmetry (the tilt plane) as 
illustrated in Fig. 4b. By analogy to the invariance of the direction of 
pseudovectors under mirror reflection, the indigenous polarity of the \textit{Sm-C} 
layers can then be termed as pseudovector-type.

In order to quantify the indigenous pseudovector polarity of the \textit{Sm-C} layers and 
to illustrate its molecular origins we consider an idealised molecular model 
embodying the minimal asymmetry required for the appearance of tilted 
orientational ordering and positional disorder within the layers. For 
simplicity we consider a rigid molecule whose shape has the same symmetries 
as the \textit{Sm-C} phase (equivalently, the minimal asymmetry compatible with the 
\textit{Sm-C} phase). A convenient example is the oblique cylinder of Fig. 5a. It has a 
plane of symmetry, a twofold axis and an inversion centre. Objects of this 
shape have been used as molecular models in a theory of the \textit{Sm-C} phase by Somoza 
and Tarazona [20]. In the present context they are 
used merely in order to convey the molecular symmetry by means of simple 
images. 
\begin{figure}[htbp]
\centerline{\includegraphics[width=3.01in,height=2.07in]{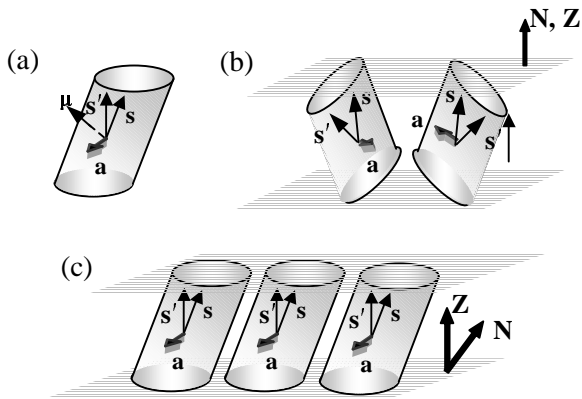}}
\caption{(a) Oblique cylinder idealization of a molecular structure bearing 
the minimal asymmetry $(C_{2h} )$ required for the formation of the Sm-C 
phase. The vectors ${\rm {\bf s}},{\rm {\bf {s}'}}$ are on the 
oblique-cylinder plane of symmetry, to which the pseudovector ${\rm {\bf a}} 
= {\rm {\bf s}}\times {\rm {\bf {s}'}}$ is perpendicular. The possible 
presence of a dipole moment ${\rm {\bm \mu }}$ directed out of the ${\rm 
{\bf s}},{\rm {\bf {s}'}}$ plane would render the structure chiral. (b) 
Opposite orientations of a are equally probable in the Sm-A phase. (c) The 
tilted molecular arrangement in the Sm-C layer favours one of the two 
opposite orientations of a perpendicular to the tilt plane, thus generating 
psedovector indigenous polarity.}
\label{fig5}
\end{figure}

Let the vectors \textbf{s} and ${\rm {\bf {s}'}}$ be directed along the 
cylinder axis and normal to the oblique surfaces respectively. The two 
halves on either side of the symmetry plane of the molecule (the two 
``faces'' of the molecule, for brevity) are distinct mirror images of each 
other. The pseudovector ${\rm {\bf a}} = {\rm {\bf s}}\times {\rm {\bf 
{s}'}}$ can then be used for the distinction between the two faces. In the 
\textit{Sm-A} phase there is clearly nothing to make any given orientation of ${\rm {\bf 
a}}$ have a different probability from its opposite (Fig. 5 b) whereas the 
tilted ordering of the \textit{Sm-C} phase induces a preference of one of the two 
opposite orientations perpendicular to the tilt plane over the other (Fig. 
5c). This preference gives rise to a nonvanishing value of the ensemble 
average $\left\langle {\rm {\bf a}} \right\rangle $ and allows the 
definition of the (pseudovector) order parameter ${\rm {\bf P}}_{\rm {\bf 
I}} = \left\langle {\rm {\bf a}} \right\rangle / \left| {\rm {\bf a}} 
\right|$ that quantifies the indigenous polarity. Clearly, the indigenous 
polarity is not related to chirality. This is depicted in Fig. 4b where it 
is apparent that the presence of pseudovector ${\rm {\bf P}}_{\rm {\bf I}} $ 
does not differentiate a layer from its mirror image. However, if the 
indigenously polar \textit{Sm-C} phase is to exhibit spontaneous electric polarisation, 
it is necessity that the molecules possess an electric dipole moment with a 
nonvanishing component along ${\rm {\bf a}}$. In that case the indigenous 
polarity ${\rm {\bf P}}_{\rm {\bf I}} $ will generate a spontaneous electric 
polarisation ${\rm {\bf P}}_s^e $ normal to the tilt plane according to the 
relation 
\begin{equation}
\label{eq6}
{\rm {\bf P}}_s^e = {\cal N}\left\langle {\rm {\bm \mu }} \right\rangle = 
{\cal N}{\bm \mu}_ \bot ^\ast {\rm {\bf P}}_{\rm {\bf I}} \quad ,
\end{equation}
\noindent
where $\mu_ \bot ^\ast = {\rm {\bm \mu }} \cdot {\rm {\bf a}} / \left| {\rm 
{\bf a}} \right|$ is a pseudoscalar measure of the strength of what is often 
referred to as the ``transverse dipole moment''. It is obvious, however, 
that the attachment of a transverse dipole component to the molecules (see 
Fig. 5a) destroys the mirror plane symmetry of their structure, which thus 
becomes chiral. In this sense, the molecular parameter $\mu_\bot^\ast$ 
provides a measure of the electrostatic chirality of the molecular 
structure. The electrically polar layer is no longer identical to its mirror 
image (see Fig. 4c), i.e. the layers have become electrostatically chiral. 

According to the above considerations, the relevance of vector and 
pseudovector polarity to spontaneous polarisation can be summarised by 
stating that vector polarity is not necessarily present and, when present, 
can give rise to spontaneous polarisation even if the molecules are not 
chiral whereas pseudovector polarity is necessarily present in all tilted 
smectics but can give rise to electric spontaneous polarisation only if the 
molecules are chiral. This difference is born by equations (\ref{eq3}) and (\ref{eq6}), 
which are formally identical but the two different mechanisms by which ${\rm 
{\bf P}}_s^e $ is generated are explicitly reflected on

(i) the different physical significance of the molecular quantities $\mu 
_\parallel$ (scalar measure of longitudinal dipole) and $\mu _\bot^\ast$ 
(pseudoscalar measure of chirality-related transverse dipole) and 

(ii) the molecular order parameters $\left\langle {\rm {\bf m}} 
\right\rangle $ and ${\rm {\bf P}}_{\rm {\bf I}} $ quantifying respectively 
the direct (vector) polarity and the indigenous, tilt-generated, 
pseudovector polarity.

Having used an explicit molecular idealisation (Fig. 5) to illustrate the 
significance of the (pseudovector) indigenous polarity, its role in giving 
rise to spontaneous electric polarisation in tilted smectics, its 
differentiation from the latter polarisation and from the (vector) polarity 
appearing in orthogonal smectics and nematics, it is useful to give a 
generalised description of the above in terms of molecular and phase 
symmetry: The \textit{Sm-C} layers have $C_{2h}$ symmetry. The plane of symmetry is 
identified with the tilt plane and the twofold symmetry axis $C_{2}$ is 
perpendicular to that plane. As a result of the rotational symmetry about 
this axis, the projection of any vector or pseudovector order parameter onto 
the tilt plane necessarily vanishes. Furthermore, the plane of symmetry 
makes all vector order parameter components along the $C_{2 }$ axis vanish 
but allows the appearance of pseudovector components along that axis. To 
exclude the possibility of such pseudovector components would require at 
least one symmetry axis in the tilt plane, but such symmetry axis is of 
course precluded by the monoclinic symmetry of the tilted layers. Hence, 
tilt generates pseudovector indigenous polarity ${\rm {\bf P}}_{\rm {\bf I}} 
$ along the $C_{2 }$axis of the \textit{Sm-C} layers. To identify the molecular origin of 
this polarity it is necessary to consider molecular symmetry. To this end, 
it is sufficient to consider molecules that are not less symmetric (in the 
statistical sense, when referring to rigid molecules) than the phase itself, 
i.e molecules with $C_{2h}$ symmetry. Again, the existence of a plane of 
symmetry together with the lack of any symmetry axes on that plane makes it 
possible to define only molecular pseudovectors directed perpendicular to 
the plane of symmetry, i.e. along the molecular $C_{2}$ axis. The ensemble 
average of the projection of such a pseudovector of unit length along the 
$C_{2}$ axis of the layers defines the order parameter ${\rm {\bf P}}_{\rm 
{\bf I}} $. The presence of the molecular $C_{2}$ axis is not essential to 
these considerations; only the molecular plane of symmetry is necessary in 
order to exclude molecular chirality. Now, if chirality is allowed for in 
the smectic layers then there will be no symmetry plane. The chiral analogue 
of the smectc-C phase, the \textit{Sm-C*}, has only one $C_{2}$ axis. This precludes the 
appearance of any vector or pseudovector order parameter components in the 
tilt plane (defined as the plane perpendicular to the $C_{2}$ axis, but no 
longer a symmetry plane). It allows, however, the appearance of both vector 
and pseudovector components in the $C_{2}$ direction. This leads to the 
possibility of simultaneous (pseudovector) indigenous polarity and (vector) 
spontaneous polarisation in chiral tilted smectics. 

Combinations of pseudovector polarity with the vector polarity described in 
the previous section for orthogonal smectics are possible and generate 
several polar variants of the \textit{Sm-C} phase. Thus, longitudinal vector polarity is 
present, in addition to indigenous pseudovector polarity, in the tilted 
analogues \textit{Sm-C}$_{2}$, \textit{Sm-C}$_{d}$ and \textit{Sm-}$\tilde {C}$ (Fig. 6 a,b,c) of the orthogonal 
bilayer variants of the \textit{Sm-A}. The tilted sublayers in these phases are polar 
along the director, i.e. have broken ${\rm {\bf N}} \Leftrightarrow - {\rm 
{\bf N}}$ symmetry, and are therefore lacking the twofold symmetry axis 
(normal to the tilt plane) of the common \textit{Sm-C} layers. However, this symmetry is 
restored in the bilayers due to the opposite longitudinal polarity 
$\left\langle {\rm {\bf m}} \right\rangle$ of their constituent sublayers. 
As in the case of orthogonal smectics, these sublayers could exhibit 
spontaneous electric polarisation ${\rm {\bf P}}_s^e$ along the director 
(consequently in the tilt plane and tilted with respect to the layer 
normal). For chiral molecules with transverse dipole moments this component 
of ${\rm {\bf P}}_s^e $ would be superposed to the component (perpendicular 
to the tilt plane) associated with the indigenous polarity and thus the 
total ${\rm {\bf P}}_s^e $ in each sublayer would be in the plane containing 
the director ${\rm {\bf N}}$ and the tilt pseudovector ${\rm {\bf t}}$. 

\begin{figure}[htbp]
\centerline{\includegraphics[width=3.4in]{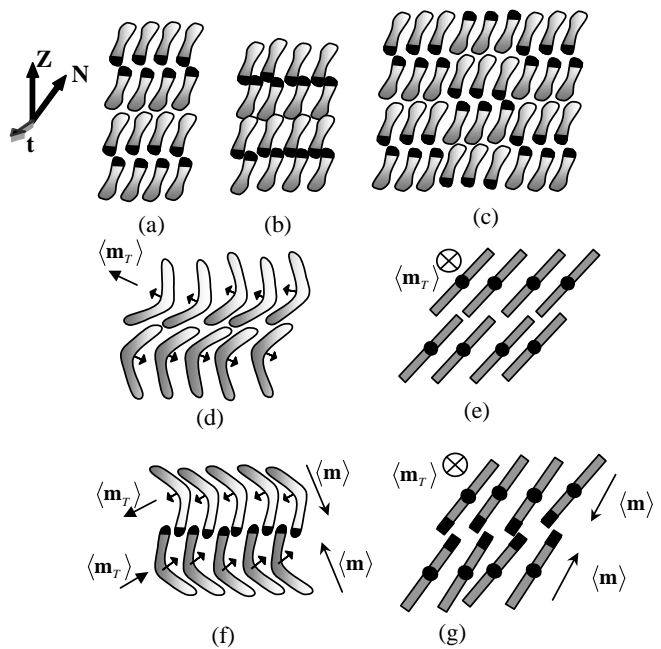}}
\caption{Tilted analogues of the polar smectics of figures (3a,b,c,d,f). In 
(e ) and (g) the transverse vector polarity order parameter $\left\langle 
{{\rm {\bf m}}_T } \right\rangle$ is directed normal to the to the tilt 
plane.
}
\label{fig6}
\end{figure}

Transverse vector polarity can be combined with indigenous pseudovector 
polarity in many ways and results in the various banana-type tilted 
smectics. Thus, the vector order parameter $\left\langle {{\rm {\bf m}}_T } 
\right\rangle $ can be in the tilt plane (therefore perpendicular to the 
direction of the indigenous polarity ${\rm {\bf P}}_{\rm {\bf I}} )$, 
resulting in the layer structure of the $C_{B1}$ phase (Fig. 6d), or it can 
be perpendicular to the tilt plane (therefore parallel to ${\rm {\bf 
P}}_{\rm {\bf I}} )$, resulting in $C_{B2}$ layers Fig. 6e, or it could be in 
any intermediate direction (within the plane perpendicular to the director 
and to the tilt plane) resulting in $C_{G}$ layers 
[21]. Finally, the possible polymorphism of polar 
tilted smectics could be further enriched by combining the indigenous 
pseudovector polarity ${\rm {\bf P}}_{\rm {\bf I}} $, simultaneously with 
the transverse, $\left\langle {{\rm {\bf m}}_T } \right\rangle $, and the 
longitudinal, $\left\langle {\rm {\bf m}} \right\rangle $, vector-type 
polarities (Fig. 6f, g).

\section{POLARITY IN COLUMNAR PHASES }

Polar ordering, both of the vector and pseudovector type, is possible in 
columnar liquid crystals [22]. By analogy to smectic 
layers, the columns in orthogonal columnar phases can exhibit longitudinal 
vector polarity, with the order parameter $\left\langle {\rm {\bf m}} 
\right\rangle $ directed along the column axis (Fig. 7a, b), or transverse 
vector polarity, with the vector order parameter $\left\langle {{\rm {\bf 
m}}_T } \right\rangle $ directed perpendicular to the column axis (Fig. 7c). 
For flat plate-like molecules, the longitudinal polarity can be generated by 
polar intermolecular interactions along the plate axis (Fig. 7a), including 
electric dipole interactions [8]. The latter 
possibility obtains because, as shown in Fig. 1c, the dipoles in 
head-to-tail configuration are separated by the ``thin'' dimension of the 
plate and thus the energy of that configuration becomes much lower than that 
of side-by-side. For bowl-shaped molecules (Fig. 7b), their polar ordering 
in each column is driven by their shape-dictated directional stacking.

Transverse polarity can be generated in columns by amphiphilic interactions 
resulting from the partitioning of the plate into sectors of different 
philicity as shown in Fig. 7c. In this case the up-down symmetry of the 
column is maintained but the rotational symmetry about the column axis is 
lost and the thus columns, by becoming transversely polar, become inevitably 
biaxial. 

\begin{figure}[htbp]
\centerline{\includegraphics[width=3.15in,height=1.71in]{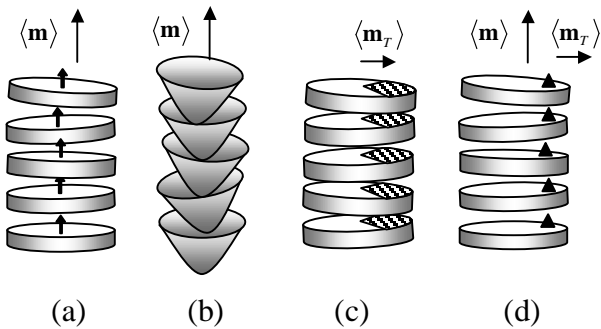}}
\caption{Schematic representation of molecular arrangements in polar columns 
of some orthogonal columnar phases. (a) Longitudinally polar discotic 
column. (b) Longitudinally polar bowl-column. (c) Transversely polar 
discotic column. (d) Discotic column with longitudinal and transverse 
polarity.}
\label{fig7}
\end{figure}

Columns exhibiting both longitudinal and transverse polarity have neither 
up-down symmetry nor rotational symmetry about the column axis and could be 
generated by combining the interactions corresponding to Fig. 7a or 7b with 
those of Fig. 7c, or, for example, by off-centre polar interactions directed 
across the plates, as shown in Fig. 7d. 

In tilted columnar phases [23] the director ${\rm 
{\bf N}}$, i.e. the direction of alignment of the molecular plate-normals, 
does not coincide with the direction ${\rm {\bf Z}}$ of the axis of the 
column. These two distinct directions define the tilt plane of the column. 
The respective tilt pseudovector ${\rm {\bf t}}$ is defined in terms of 
${\rm {\bf N}}$ and ${\rm {\bf Z}}$ as in Eq. (\ref{eq6}). The tilted ordering of 
the columns inflicts a pseudovector type polarity on the structure. The 
molecular origin of the indigenous pseudovector polarity for tilted columnar 
liquid crystals can be illustrated in close analogy with that of tilted 
smectics [24] by considering idealised molecules of 
oblique disc shape as in Fig. 8a. The two molecular vectors ${\rm {\bf s}}$ 
and ${\rm {\bf {s}'}}$ define a pseudovector ${\rm {\bf a}} = {\rm {\bf 
s}}\times {\rm {\bf s}}'$ whose direction differentiates between the two 
mirror image halves, or ``faces'', of the oblique disc and whose ensemble 
average $\left\langle {\rm {\bf a}} \right\rangle $ in the tilted column 
defines the indigenous polarity order parameter ${\rm {\bf P}}_{\rm {\bf I}} 
= \left\langle {\rm {\bf a}} \right\rangle / \left| {\rm {\bf a}} \right|$ 
(Fig 8c). The asymmetry is removed in an orthogonal column (Fig. 8b) making 
$\left\langle {\rm {\bf a}} \right\rangle = 0$. The direction of ${\rm {\bf 
P}}_{\rm {\bf I}} $ is along the tilt pseudovector ${\rm {\bf t}}$, i.e. 
perpendicular to the tilt plane of the column, and is therefore compatible 
with the two-fold rotational symmetry about ${\rm {\bf t}}$. Moreover, the 
pseudovector character of ${\rm {\bf P}}_{\rm {\bf I}} $ makes it compatible 
with the mirror plane symmetry of the tilted columns. As in the case of 
tilted smectics, pseudovector polarity in tilted columnars can be combined 
with longitudinal vector polarity (along the director) or with transverse 
vector polarity, or with both, to provide polar columns in which either the 
mirror plane symmetry or the twofold rotation symmetry or both symmetries 
are broken. 

\begin{figure}[htbp]
\centerline{\includegraphics[width=3.1in]{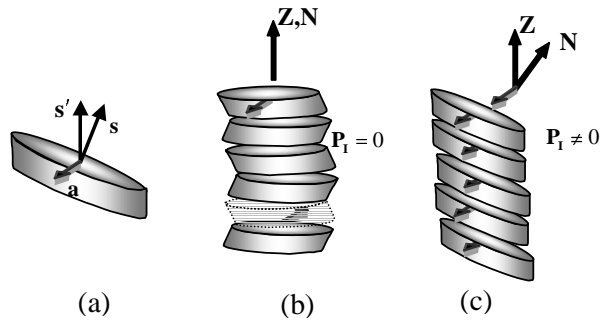}}
\caption{Oblique disc idealization of a molecular structure bearing the 
minimal asymmetry required for the formation of tilted columnar phases. The 
vectors ${\rm {\bf s,{s}'}}$ are on the oblique disc plane of symmetry, to 
which the pseudovector ${\rm {\bf a}} = {\rm {\bf s}}\times {\rm {\bf 
{s}'}}$ is perpendicular. (b) Opposite orientations of \textbf{a} are 
equally probable in an orthogonal column. (c) The accommodation of the 
oblique disc in a tilted column favours one of the two opposite orientations 
of \textbf{a} perpendicular to the tilt plane, thus generating psedovector 
indigenous polarity.} 
\label{fig8}
\end{figure}

\section{CONCLUSIONS AND DISCUSSION}

Strictly, none of the known low molar mass liquid crystals exhibit, in their 
thermodynamically equilibrated bulk state, spontaneous macroscopic 
polarisation, electric or magnetic. The repetitive building blocks (layers, 
columns, etc) of phases with partial positional order are often polar but 
arranged in polarity neutralising patterns (bilayer, antiferroelectric, 
helical, etc) to produce macroscopically a-polar phases. In this paper we 
have considered the different types of polar self-assembly of the building 
blocks, we have quantified the resulting polar ordering in terms of 
molecular order parameters and we have identified the underlying molecular 
symmetries and intermolecular interactions.

Vector-type and pseudovector-type polarity refer to two distinct asymmetries 
of molecular ordering. Vector type applies when there is no plane of 
symmetry and no symmetry axis perpendicular to the direction of polarity. 
Pseudovector-type applies when there is no symmetry axis perpendicular to 
the direction of polarity but there is a plane of symmetry. Vector polarity 
is theoretically possible for nematics but requires interactions of rather 
unusual position-orientation dependence. In contrast, amphiphilic, steric 
and electrostatic dipole interactions can, separately or cooperatively, give 
rise to vector polarity in the repetitive units of smectic and columnar 
phases. The polarity in these cases could appear along the primary direction 
of alignment (longitudinal) or in a direction perpendicular to it 
(transverse) or both. Pseudovector polarity is produced by the tilted 
molecular arrangement in all a-chiral tilted smectic and columnar phases. It 
is driven by the same interactions that give rise to the tilted ordering and 
reflects the asymmetry imposed on the molecular orientations by the packing 
constraints. 

Introducing polarity, with its different types and combinations, to the 
common a-polar modes of molecular self-organisation enriches the 
polymorphism of mesophases with many polar variants. A good understanding of 
the molecular asymmetries and interactions that favour particular types of 
polar ordering is certainly essential to the design of such materials. 
However, in addition to these qualitative assessments, quantitative 
estimates could be crucial to the successful design. This is so because the 
thermodynamic stability of mesophases usually rests on a delicate balance 
between order-promoting interactions that are counteracted by entropy 
lowering. Therefore, the interactions need to be strong enough to produce 
the required molecular correlations but not so strong as to destabilise 
liquid crystallinity in favour of a solid phase. At present, theory and 
computer simulations could provide useful quantitative input but there are 
also many complications in obtaining realistic estimates. For example, 
almost all liquid crystal forming molecules are flexible. For qualitative 
considerations molecular symmetry is understood to refer to the 
``conformationally averaged'' molecule and so are the intermolecular 
interactions. In a quantitative treatment, however, the individual molecular 
conformations (normally lacking any element of symmetry) have to be 
considered together with the detailed interactions of the various molecular 
segments. This usually introduces an enormous number of degrees of freedom. 
Assuming that the latter can be handled computationally, the predictive 
quality of the results rests on the detailed modelling of the intra- and 
inter-molecular interactions. A reasonably realistic modelling of 
interactions in liquid crystals is in general very difficult due to the 
subtlety of these interactions. For example, interactions associated with 
partial charge distributions are routinely modelled in terms of fixed 
electric dipole moments, quadrupole moments etc. This kind of crude 
modelling does not allow for the possibly significant deformations of the 
charge distribution caused by intermolecular interactions (polarisability). 
It also disregards the fact that the multipole expansion of a charge 
distribution is valid only at distances that are large compared to the 
spatial extent of the distribution. As a result the respective quantitative 
predictions would be at best indicative of the possible behaviour of the 
real system.
\section*{Acknowledgements}
This paper is dedicated to the memory of Pier Luigi Nordio.

Partial financial support for this work came from the EU \textit{TMR} (contract 
FMRX-CT97-0121) and through the funding of the Polymer Science and 
Technology Graduate Studies Programme of the University of Patras. 


\begin{thebibliography}{99}

\bibitem{1.}Recent reviews can be found in B. Groh and S. Dietrich, \textit{New approaches to problems in liquid state theory}, (Kluwer, 
1999), pp. 173-196; S.T. Lagerwall, \textit{J. Phys: Condens. Mat}. \textbf{8}, 9143 (1996); P.I.C. 
Teixeira, J.M. Tavares and M.M.T. da Gama, \textit{J. Phys: Condens. Mat.} \textbf{12}, R411, (2000); S.T. 
Lagerwall, \textit{Ferroelectric and Antiferroelectric Liquid Crystals}, (John Wiley {\&} Sons, NY, 1999); I. Musevic, R. Blinc and B. 
Zeks, \textit{The Physics of Ferroelectric and Antiferroelectric Liquid Crystals, }(World Scientific, 2000).

\bibitem{2.} Technically, liquid metals exhibiting ferromagnetism in the undercooled 
state (T. Albrecht, C. Buhrer, M. Fahnle, K. Maier, D. Platzek, and J. 
Reske, \textit{Appl. Phys.} A \textbf{65}, 215 (1997); A.N. Grigorenko P.I. Nikitin, A.Y. 
Toporov, A.M. Ghorbanzadeh, A. Perrone, A. Zocco and M.L. De Giorgi, \textit{Appl. Phys. Lett}. 
\textbf{72}, 3455(1998)) could be termed as polar nematics but these systems 
are otherwise quite different form what is commonly considered as low molar 
mass liquid crystals.

\bibitem{3.} D.J. Photinos, CD Poon, E.T. Samulski and H. Torium, \textit{J. Phys. Chem.} \textbf{96}, 8176 
(1992). 

\bibitem{4.} Historically, dipolar interactions where the first to be considered as 
the underlying interaction for the formation of mesophases: M. Born, \textit{Sitz. Phys. Math.} 
\textbf{25}, 614 (1916); M. Born, \textit{Ann. Phys.} \textbf{55}, 221 (1918). 

\bibitem{5.} M.E. van Leeuwen and B. Smit, \textit{Phys. Rev. Lett.} \textbf{71}, 3991 (1993).

\bibitem{6.} R.P. Sear, \textit{Phys. Rev. Lett.} \textbf{76}, 2310 (1996).

\bibitem{7.} S.C. McGrother, A. Gil-Villega and G. Jackson, \textit{J. Phys: Condens. Mat. }\textbf{8}, 9649 (1996).

\bibitem{8.} J. J. Weis, D. Levesque, and G. J. Zarragoicoechea,\textit{ Phys. Rev. Lett.} \textbf{69}, 913 
(1993).

\bibitem{9.} C. Tschierske, \textit{J. Mater. Chem.} \textbf{11}, 2647 (2001).

\bibitem{10.} R. Berardi, M. Ricci and C. Zannoni\textit{, Chem. Phys. Chem}. \textbf{2}, 443 (2001).

\bibitem{11.} J.W. Goodby,\textit{ Handbook of Liquid Crystals: Low Molecular Weight Liquid Crystals I: Calamitic Liquid Crystals, }edited by D. Demus, J.W. Goodby, G.W. Gray, H.W. Spiess and 
V. Vill, (Wiley-Vch, NY, 1998), and references therein.

\bibitem{12.} R. Berardi, S. Orlandi, D. J. Photinos, A. G. Vanakaras and C. Zannoni, 
\textit{Phys. Chem. Chem. Phys.} \textbf{4}, 770 (2002).

\bibitem{13.} P.J. Photinos and A. Saupe, \textit{Phys. Rev. A} \textbf{13}, 1926 (1975).

\bibitem{14.} A. Eremin, S. Diele, G. Pelzl, H. Nadasi, W. Weissflog, J. Salfetnikova 
and H. Kresse, \textit{Phys. Rev. E } \textbf{64}, 051707 (2001).

\bibitem{15.} D. Levesque, J. J. Weis, and G. J. Zarragoicoechea, \textit{Phys. Rev. E} \textbf{47}, 496 
(1993).

\bibitem{16.} A. Gil-Villegas, S.C. McGrother and G. Jackson \textit{Chem. Phys. Lett.}, \textbf{269}, 441 
(1997).

\bibitem{17.} P.-G. de Gennes, \textit{The Physics of Liquid Crystals} (Clarendon Press, Oxford, 1974).

\bibitem{18.} P.K. Karahaliou, A.G. Vanakaras and D.J. Photinos, \textit{Phys. Rev. E} \textbf{65}, 031712 
(2002).

\bibitem{19.} D.J. Photinos and E.T. Samulski, \textit{Science} \textbf{270}, 783 (1995).

\bibitem{20.} A.M. Somoza and P. Tarazona, \textit{Phys. Rev. Lett.}, \textbf{61}, 2566 (1988).

\bibitem{21.} P.E. Cladis, H.R. Brand, and H. Pleiner, \textit{Liq. Cryst. Today} \textbf{9}, 3/4 (1999).

\bibitem{22.} D. Demus, J.W. Goodby, G.W. Gray, H.W. Spiess and V. Vill (editors),\textit{ Handbook of Liquid Crystals: Low Molecular Weight Liquid Crystals II: Discotic and Non-Conventional Liquid Crystals} 
(Wiley-Vch, NY, 1998), and references therein.

\bibitem{23.} H. Bock, W. Helfrich, \textit{Liquid Crystals} \textbf{18}, 378 (1995).

\bibitem{24.} A.G. Vanakaras, D.J. Photinos, E.T. Samulski, \textit{Phys. Rev E} \textbf{57}, R4875 
(1998).


\end{thebibliography}
\end{document}